\documentclass[letterpaper,12pt]{article}
\usepackage{fullpage}
\usepackage{times}
\usepackage{amsmath}
\usepackage{graphicx}
\usepackage{cite}

\renewcommand{\baselinestretch}{1.5}

\begin{document}

\noindent {\LARGE \bf \sf A simple monatomic ideal glass former: the
  glass transition by a first-order phase transition above the melting
  point}

\vskip 0.5cm

\noindent
M\aa ns Elenius$^1$, Tomas Oppelstrup$^{1,3}$ and Mikhail Dzugutov$^2$

\vskip 0.5cm

\noindent
{\it $^1$Department of Numerical Analysis and ~$^2$Department of
Materials Science and Engineering, Royal Institute of Technology,
100 44 Stockholm, Sweden\\
$^3$Lawrence Livermore National Laboratory, Livermore, California 94551, USA}

\vskip 0.5cm

\noindent

\abstract{A liquid can form under cooling a glassy state either as a
  result of a continuous slowing down or by a first order
  polyamorphous phase transition. The second scenario has so far
  always been observed below the melting point where it interfered
  with crystalline nucleation. We report the first observation of the
  liquid-glass transition by a first order phase transition above the
  melting point.  The observation was made in a molecular dynamics
  simulation of a one-component system with a model metallic pair
  potential. This is also the first observation of a simple monatomic
  ideal glass former -- a liquid that avoids crystallization at any
  cooling rate. Besides its conceptual importance, this result
  indicates a possibility of existence of metallic ideal glass
  formers.}

\newpage Periodic phases are commonly thought to be thermodynamically
favoured at sufficiently low temperatures by all substances. The
relaxation time of a liquid in its domain of thermodynamic stability,
above $T_m$, is usually much smaller than the value of $10^2 s$ that
defines the glass-transition temperature $T_g$
\cite{debenedetti}. Therefore, a fundamental problem of the glass
science is to avoid the interference of crystalline nucleation when
cooling a liquid within the metastability domain $T_g< T < T_m$. The
ideal glass former\cite{angell08} is defined by the condition
$T_g>T_m$ which excludes the possibility of its crystallisation upon
cooling. This class of glass formers has so far been found to include
atactic polymers \cite{ kruger } and some aqueous solutions of
electrolytes \cite{angellsare}. A simple monatomic liquid that does
not crystallize upon cooling has so far never been found. The question
of its existence is intellectually challenging, conceptually important
and also of a significant technological interest, particularly for the
area of metallic glasses.

A straightforward strategy for reducing $T_m$ relative to $T_g$
is to frustrate the crystallization by structural complexity. In
non-polymeric systems, this can be achieved by tuning anisotropic
interaction \cite{molinero}, by a judiciously designed pair
potential\cite{dzugutov02}, or by composing multicomponent
eutectic mixtures \cite{angellsare, johnson99,
schneider}. Crystallization of a liquid upon cooling can also be
precluded by a direct first order polyamorphous transition to a
glassy state. Such a transition, however, has so far always been
observed below $T_m$ \cite{sastry03}, where the presence of a
liquid-liquid spinodal significantly enhances crystalline
nucleation \cite{frenkel,oxtoby} . Moreover, a liquid-liquid
transition has never been observed in a monatomic system upon
cooling above $T_m$ \cite{sciortino05}. The Jagla soft-core pair
potential produces two stable liquid phases. However, because of
a significant difference in densities, the transition between
these phases has only been found under compression and not upon
cooling \cite{limei06, limei09}.

Here, we report a simple monatomic liquid that remains stable
with respect to crystallization upon cooling but performs a first
order polyamorphous phase transition to a phase with a
mesoscopic-range order and the rate of structural relaxation
characteristic of the glassy state. Thus, the liquid, avoiding
crystallization at any cooling rate, is an ideal glass former.

The liquid was explored in a molecular dynamics simulation. It
used a model pair potential \cite{doye} (named Z2 in that
reference) designed to imitate effective interionic interaction
potentials in liquid metals with characteristic Friedel
oscillations \cite{moriarty, mermin}. The simulation was
performed using a system of $N=128000$ particles. A smaller
system of $N=3456$ particles has also been explored in order to
test the size-dependence of the observed phenomenon.  We note
that the main repulsive part of the pair potential used in this
model closely approximates that of the Lennard-Jones
potential. Interpreting the latter as a model potential of argon
\cite{hansen}, the units of length and time used in this
simulation correspond to 0.34 nm and $2.16$ ps, respectively. All
other quantities reported here are expressed in these reduced
units.

We cooled the large system from the high-temperature liquid state
isochorically at the number density $\rho=0.85$ by changing the
temperature $T$ in a stepwise manner, with a comprehensive
equilibration at each temperature step.  The results presented in
Fig. 1a show a discontinuous enthalpy drop on cooling at
$T=0.72$. Reheating of the lower temperature results in a pronounced
hysteresis, an unambiguous signature of a first order phase
transition.

Fig. 1b shows the structure factor $S(Q)$ for the two phases. The
similarity of the two curves indicates that we observe a first order
transition between two similarly structured liquid phases, to be
referred to as the high-temperature liquid (HTL) and the low
temperature liquid (LTL). The split in second peak of $S(Q)$ featured
by both liquid phases indicates a strong tetrahedral local order
\cite{sadoc}.

We now explore the transition domain in the $P-\rho$ phase diagram at
constant $T$. Fig. 1c shows the LTL isotherms produced by heating,
compression and/or expansion of the original states of that phase
obtained under isochoric cooling at $\rho=0.85$. All the data were
produced after comprehensive equilibration. The regions of infinite
compressibility indicate the existence of a domain of spinodal
instability interposed between the two liquid phases. We find that the
phase behaviour of the small system (N=3456) reproduces the behavior
of the large system along the $T=0.75$ isotherm.  Expectedly, the
small system size results in shrinking of the instability domain.

Infinite compressibility in the spinodal domain manifests in diverging
long-wavelength limit of $S(Q)$, a measure of the density fluctuations
on the system-size scale. In a macroscopic system, $S(0)$ is related
to the isothermal compressibility $\chi_T$ by the compressibility
equation: $S(0)= \rho k_B T \chi_T$ \cite{hansen}.  Fig. 1d shows the
density-dependence of the small-$Q$ behaviour of $S(Q)$ along the
$T=0.78$ isotherm. A trend for divergence of the small-$Q$ limit of
$S(Q)$ is clearly visible at the densities where the spinodal
instability was detected from the respective isotherm in Fig. 1c. Note
that there is no divergence in the small-$Q$ limit of $S(Q)$ for the
LTL in Fig. 1b, indicating that at $T=0.68, \rho=0.85$ it is below the
spinodal domain.

Fig. 2 presents a real-space picture of the spinodal decomposition of
a system of 128000 particles along the $T=0.78$ isotherm. The plots
depict cross-sections of the coarse-grained spatial distribution of
energy.  For $\rho=0.84$ representing the HTL phase, the energy is
distributed uniformly. At $\rho=0.87$, precipitation of the LTL phase
appears as distinct large-scale low-energy domains.  Upon further
isothermal compression, the LTL domains grow and eventually percolate
as the system leaves the spinodal domain. This is indicated by the
reduction of the isothermal compressibility in the isotherms as shown
in Fig. 1d. Respectively, the HTL domains shrink and become
disconnected, arguably giving rise to the low-$Q$ pre-peak of the
respective $S(Q)$ visible in Fig. 1d. This is consistent with the
conjecture \cite{ponyatovsky03} that a LTL phase is intrinsically
heterogeneous.

In order to understand the nature of the observed phase
transition, we analysed details of the structural transformation
between the two phases.  To remove thermally induced
fluctuations, the investigated liquid configurations were
subjected to the steepest descent energy minimisation. In the
structure analysis we used an earlier developed method of
identifying the tetrahedral local order \cite{dzugutov02}. All
the particles comprising the first peak of the radial
distribution function $g(r)$ were assumed to be the nearest
neighbours. The narrow and sharp shape of the peak renders this
definition of neighbours unambiguous.

We found that the HTL-LTL transition, while sustaining a strong
tetrahedral local order, reduces the number of icosahedra by about
30\%. This paradoxical observation can be understood by inspecting a
characteristic cluster found in a low-temperature domain of the LTL in
Fig. 2. The cluster, shown in Fig. 3a, can be described as a 5-fold
tetrahelical configuration composed of axially stacked pentagonal
bipyramides.  The latter's pentagonal symmetry, however, is broken by
a linear strain of non-tetrahedral defects that involves two adjacent
helical lines of atoms, as shown in Fig 3b. The creation of such a
defect reduces the energy by removing frustration inherent to the
5-fold packing of tetrahedra.  Moreover, these defects facilitate
aggregation of the tetrahelical clusters into an extended tetrahedral
network. Fig. 3c shows two clusters which share a line of helically
stacked 4-fold defects as depicted in Fig. 3d. In this way, extended
tetrahedral configurations, apparently incompatible with periodic
order, form under cooling a network that accounts for the observed
immense viscosity of the LTL. Their inability to uniformly fill the
space renders the LTL intrinsically heterogeneous.

We now analyse the LTL dynamics at $\rho=0.85$. The diffusivity plot
in Fig. 4a, doesn't significantly change upon the transition, which is
in contrast to the more than 2 orders of magnitude diffusivity drop
observed at the glass transition in silicon \cite{sastry03}. This can
be attributed to the LTL heterogeneity demonstrated in Fig. 2, with
possibly high diffusivity within the high-energy fluid domains. The
temperature variation of the LTL diffusivity is perfectly exponential,
indicating that the HTL-LTL transition results in a fragile-to strong
crossover \cite{sastry03, saika}. We also found the LTL diffusivity to
be independent of the system size, which allowed us to use the small
system to explore the low-$T$ diffusion within affordable computer
time.

The structural relaxation is controlled by the slowest-dissipating
component of the intermediate scattering function $F(Q,t)$ (see
Methods) within the structurally relevant range of $Q$. Estimating the
$Q$-dependent relaxation time of $F(Q,t)$ in a liquid \cite{hansen} as
$\tau(Q) = S(Q) / Q^2$, we expect the small-$Q$ prepeak of $S(Q)$
observed in Fig. 1b to be the slowest dissipating structural feature
of the LTL phase. This conclusion is confirmed in the inset of
Fig. 4b. The main panel of Fig. 4b shows the time variation of
$F(Q_{pp},t)$, $Q_{pp}$ being the position of the prepeak, for
$T=0.65$ and $T=0.6$. For both temperatures, the relaxation times
clearly exceed by several orders of magnitude the time interval shown
in the plot which, in terms of argon interpretation of the reduced
units of time, corresponds to $0.65\cdot 10^{-4} s$. This makes it
possible to conclude that the LTL is a structurally arrested glassy
state.

The observed non-dissipation of $S(Q_{pp})$ implies that the spatial
distribution of the low-energy clusters in the LTL that was concluded
to give rise to the prepeak gets frozen below the spinodal domain.  To
support this conclusion, we present in Figs. 4c and 4d the evolution
of the coarse-grained potential energy distribution in the LTL at
$T=0.65$ and $T=0.60$, respectively. For each, $T$ the time interval
separating the plots corresponds to that spanned by the respective
$F(Q_{pp},t)$ shown in Fig. 4b. The apparent time invariance of the
distribution indicates that the tetrahedrally ordered low-energy
clusters like those shown in Fig. 3 form a percolating network which
remains topologically unchanged on the explored time scale.  Fig. 4e
shows a cluster composed of face-sharing tetrahedra discerned within a
low-energy domain at $T=0.65$. The percolation is reminiscent of
gelation in colloids forming a percolating network of linear clusters
composed of tetrahedra \cite{sciortino}. A similar percolation
transition was observed in the present system at low densities
\cite{elenius}. A glass transition by percolation was also found in
ideal polymeric glass-formers \cite{kruger}.

We note that the extended range of structural ordering within the
tetrahedral clusters of the LTL phase gives rise to the
anomalously high main peak of $S(Q)$, shown in Fig. 1b. This
structural peculiarity, unusual for simple liquids, can be
compared with the structure of mesophases, e.g. those formed by
proteins \cite{dobson}. Moreover, the apparent structural
heterogeneity of the LTL phase, and the relatively high
diffusivity distinguish it from the typical inorganic glasses
forming continuous tetrahedral networks. A feasible conjecture is
that these distinctions of the present glass may possibly render
it ductile and capable of self-repair, in contrast to the
well-known brittleness of the inorganic glasses. This possibility
is of significant technological interest, and deserves further
investigation.

In summary, we presented a simple one-component system with a
metal-like pair potential that behaves as an ideal glass former. Under
cooling, the system performs a polyamorphous first order phase
transition from a stable liquid phase to a glassy phase which
precludes any possibility of crystallization. The prototypical
metallic nature of the model suggests a possible new strategy for
designing bulk metallic glass formers, distinct from the currently
popular idea of frustrating crystallization by eutectically mixing a
number of atomic species with strong mismatch in size
\cite{schneider}.  As a potentially relevant observation, we mention a
strong-fragile transition found in a bulk metallic glass-forming alloy
\cite{busch2007}.

Based on these results, a possible strategy in the search for an ideal
glass former in metallic systems may be suggested. The form of the
 pair potential we used \cite{doye} is ubiquitous in interionic
effective potentials in metallic liquids\cite{moriarty}. It
incorporates the Friedel oscillations which are controlled by the
electron density\cite{mermin} and can therefore be regulated by
modifying the composition of a metallic alloy. This way of
tuning the interionic potentials in a real metallic liquid can be used  to
approximate the potential exploited in this study.

\clearpage
\renewcommand{\baselinestretch}{1.6}

\begin{figure} 
\begin{centering}
\includegraphics[width=\textwidth]{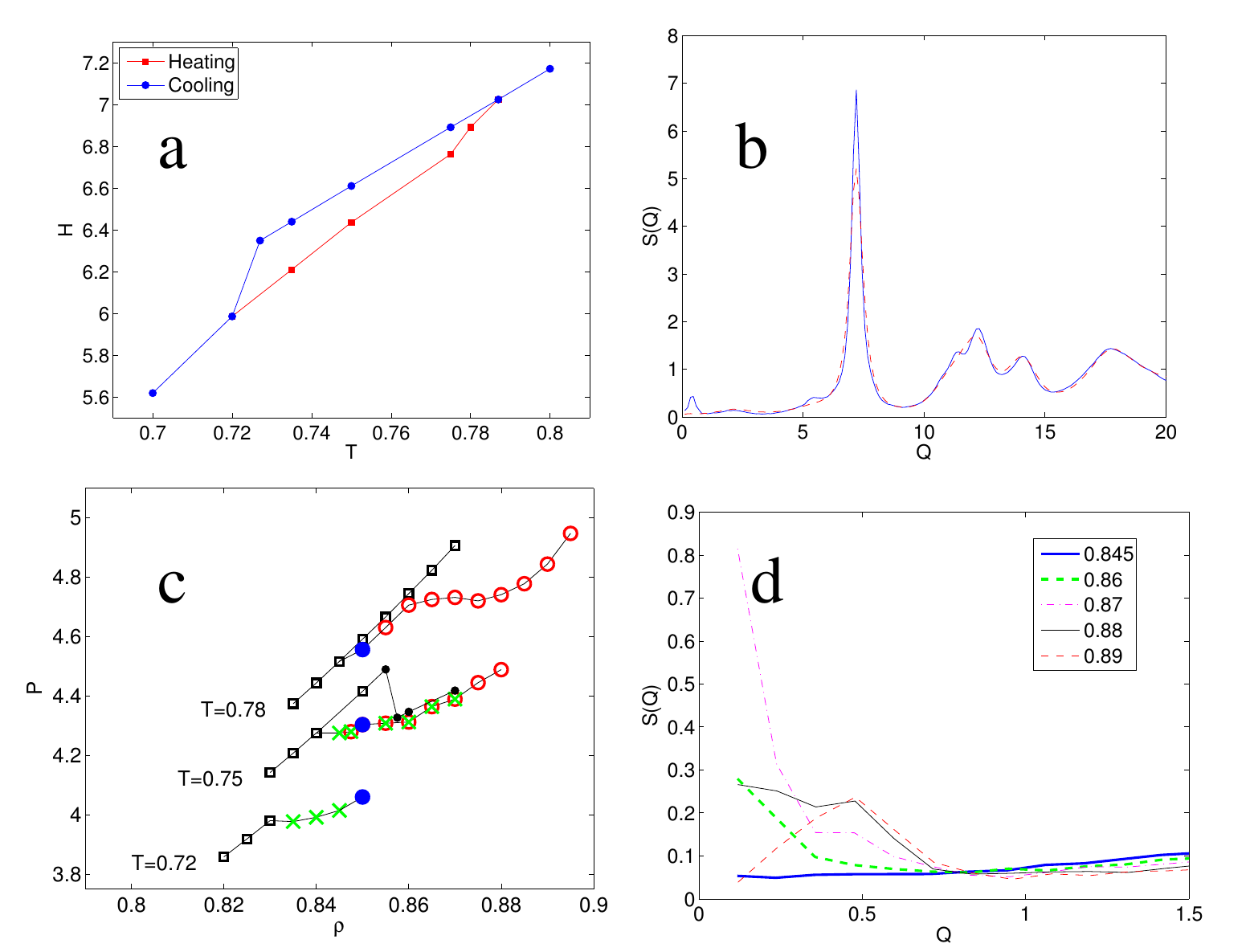}
\caption{Liquid-liquid phase transition. The data represent a system
  of $N=128000$ particles, unless indicated otherwise. {\bf
    a}. Isochoric temperature variation of the enthalpy $H$ at density
  $\rho=0.85$. Circles, cooling; squares, heating. {\bf b}. The
  structure factors $S(Q)$ of the HTL at $T=0.78$ (dashed line) and
  LTL at $T=0.68$ (solid line), both at the density $\rho=0.85$.  {\bf
    c}. Isotherms crossing the region of the liquid-liquid transition.
  Open squares: HTL. Filled circles represent the $\rho=0.85$ isochore
  shown in {\bf a}. Open circles and crosses: LTL points obtained by
  isothermal compression and expansion, respectively.  Small dots: LTL
  states simulated by the system of $N=3456$ particles. The lines are
  included as a guide to the eye.  {\bf d.} Density variation of the
  low-$Q$ behaviour of the structure factor along the $T=0.78$
  isotherm, for the indicated densities.}
\end{centering}
\end{figure}

\clearpage
\begin{figure} 
\begin{centering}
\includegraphics[width=\textwidth]{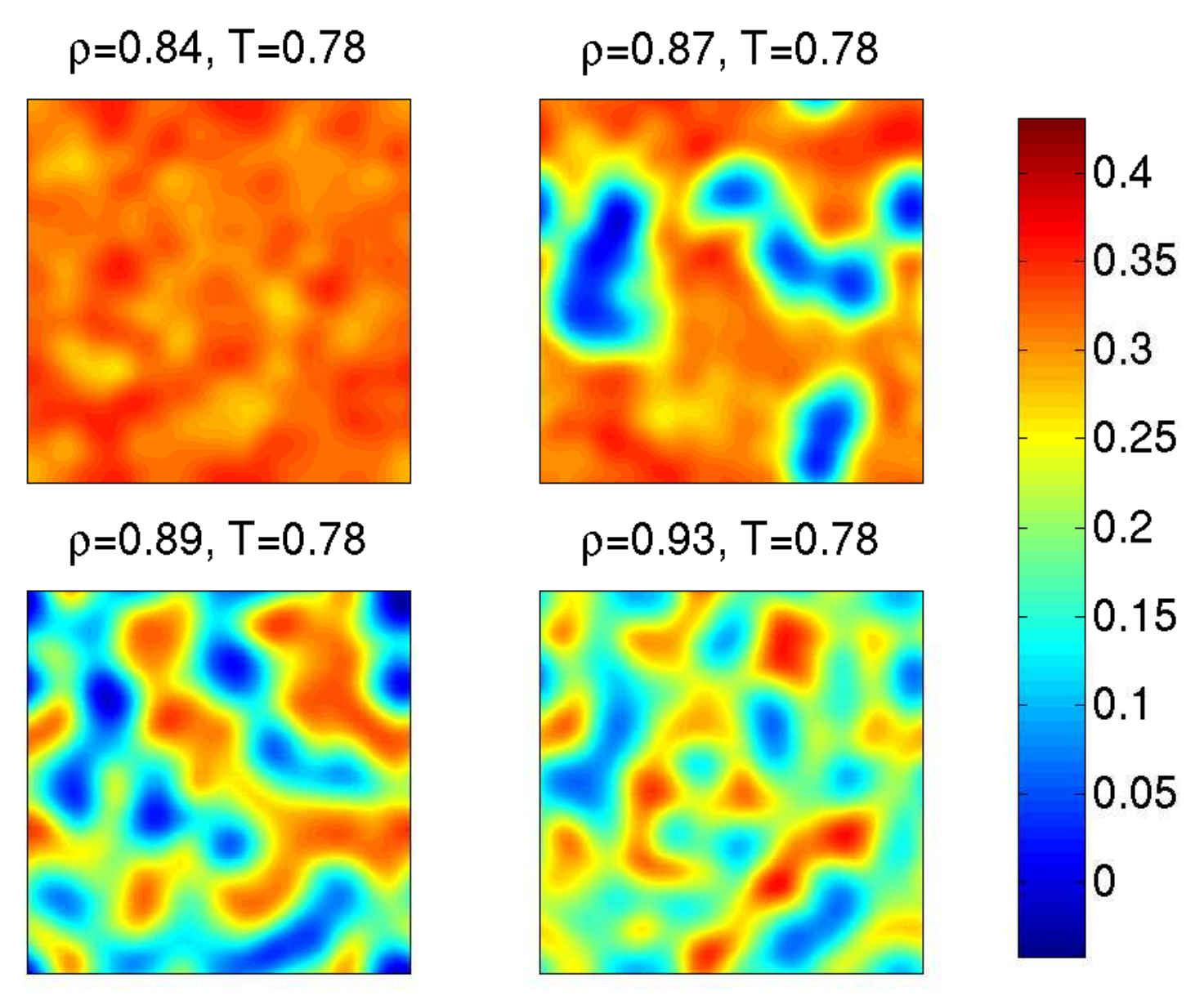}
\caption{Cross-sections of the coarse-grained energy distribution in a
  system of 128000 particles at the thermodynamic states indicated in
  the plots. Each distribution has been averaged over $10^3$ time
  steps.}
\end{centering}
\end{figure}

\clearpage
\begin{figure} 
\begin{centering}
\includegraphics[width=\textwidth]{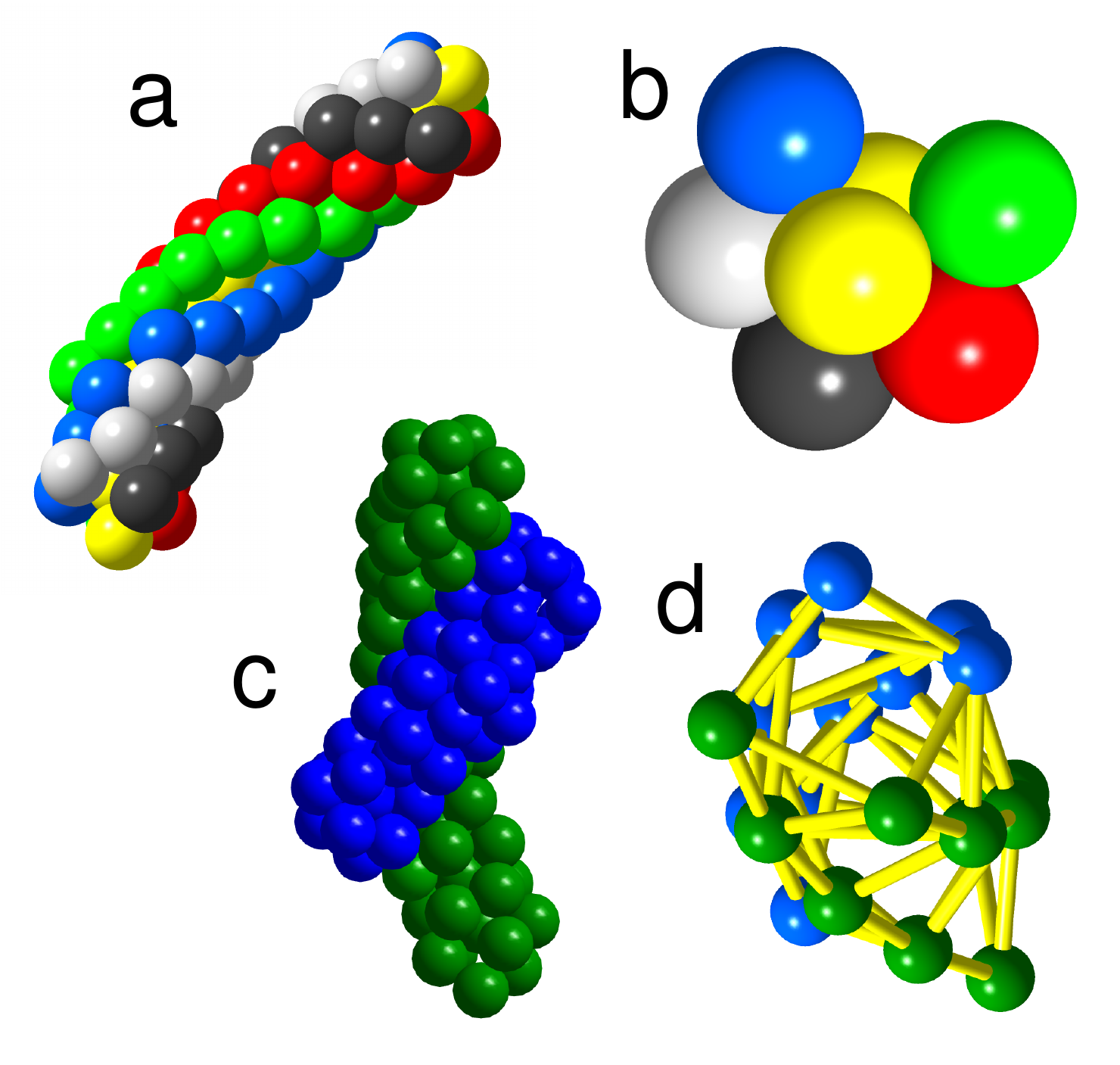}
\caption{Structure of the LTL. {\bf a}. A representative tetrahelical
  configuration discerned in a low-energy region observable in Fig. 2
  for $\rho=0.87$. The colors are a guide to the eye distinguishing
  the six constituent lines of atoms. {\bf b}. A fragment of the
  configuration shown in {\bf a} demonstrating a defect in pentagonal
  packing of tetrahedra. The colors are the same as in {\bf a}. {\bf
    c}. Aggregation of the tetrahelical cluster shown in {\bf a}
  (green) with a similar cluster (blue). {\bf d}. The linear strain of
  non-tetrahedral defects interfacing the two tetrahelical
  configurations is shown in {\bf c} using the same colors. Atoms
  fulfilling the neighbor condition as described in the text are
  connected by bonds.}
\end{centering}
\end{figure}

\clearpage
\begin{figure} 
\begin{centering}
\includegraphics[width=\textwidth]{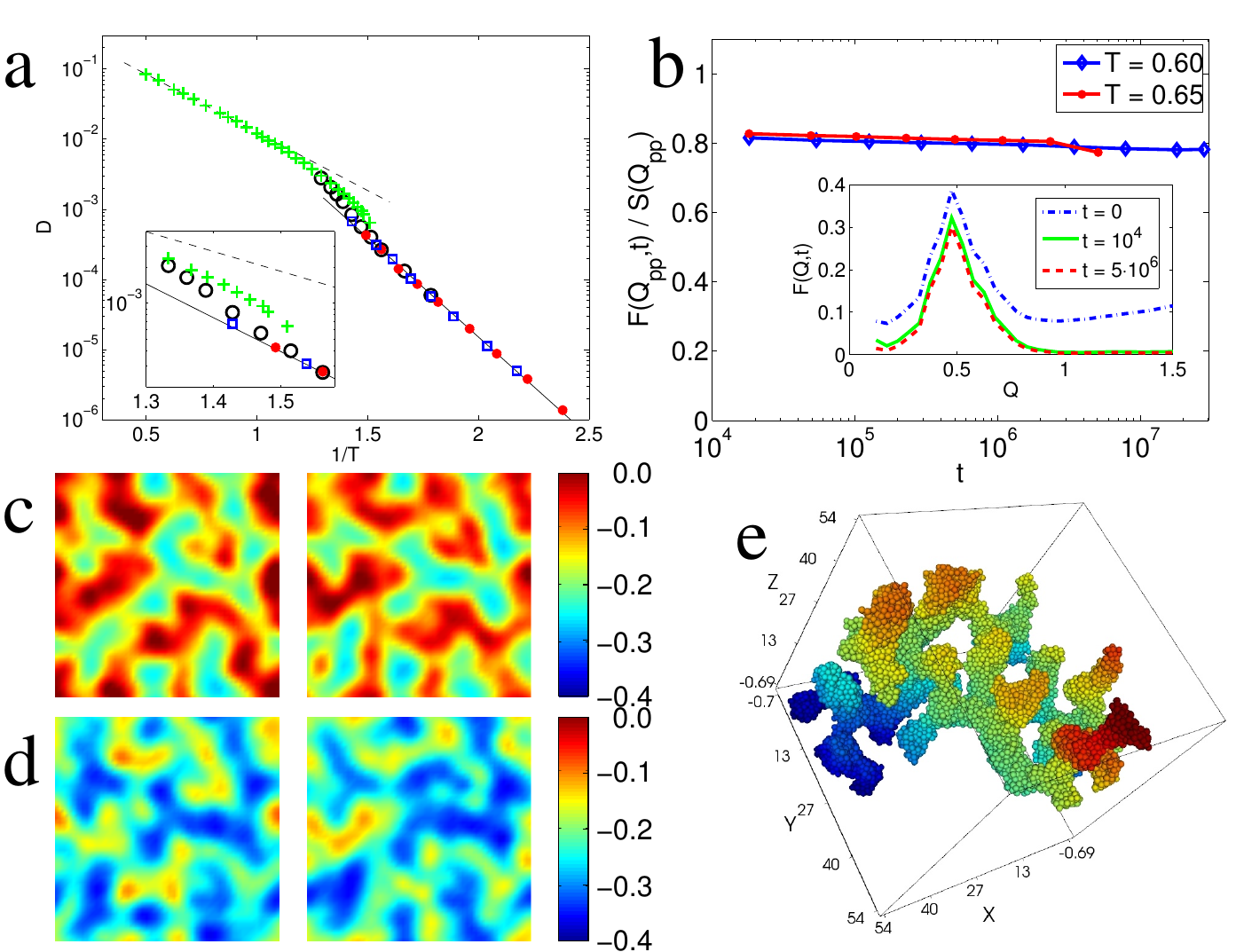}
\caption{{\bf a}. Arrhenius plot of the diffusivity at the density
  $\rho=0.85$. Crosses, HTL \cite{doye}. Open circles, LTL for the
  system of $N=128000$. particles. Dots and open squares: LTL for the
  system of $N=3456$ particles, under cooling and heating,
  respectively. The inset shows an enlargement of the transition
  area. The straight lines indicate Arrhenius fits to the data. {\bf
    b}. Main panel: the intermediate scattering function
  $F(Q_{pp},t)$, $Q_{pp}$ being the position of the low-$Q$ prepeak of
  $S(Q)$, for the LTL at $\rho=0.85$ and the two indicated
  temperatures.  Inset: the time evolution of $F(Q,t)$ in the
  small-$Q$ domain. {\bf c.}  Two cross-sections of the coarse-grained
  energy distribution at $\rho =0.85$ and $T=0.65$ separated by the
  time interval of $0.5 \cdot 10^7$.  {\bf d.} The same as in {\bf c}
  but for $T=0.6$ and the separating time interval $3 \cdot 10^7$.
  {\bf e.} A cluster composed of face-sharing tetrahedra detected
  within a low-energy domain of the configuration shown in {\bf
    c}. The variation of color indicates the variation of
  $Z$-coordinate. The cluster includes 10456 atoms.}
\end{centering}
\end{figure}


\newpage

\begin{thebibliography}{199}

\bibitem{debenedetti}
P. G. Debenedetti, F. Stillinger, 
{\it Nature} {\bf 410,} 259 (2001).

\bibitem{angell08}

C. A. Angell, 
{\it J. Non-Cryst. Solids} {\bf 354,} 4703 (2008)



\bibitem{kruger} 

  J.K. Kruger, K.-P. Bohn, R. Jimenez, and J. Schreiber, 
   {\it Colloid Polym. Sci.}, {\bf 274}, 490 (1996)

 \bibitem{angellsare} 

   E. J. Sare, and C. A. Angell, 
   {\it Journ. of Sol. Chem.}, {\bf 2}, 53 (1973)

\bibitem{molinero}

  V. Molinero, S. Sastry, and C. A. Angell, {\it Phys. Rev. Lett.}
  {\bf 97,} 075701 (2006)


\bibitem{dzugutov02}

  M. Dzugutov, S. Simdyankin, and F. Zetterling, {\it
    Phys. rev. Lett.} {\bf 89}, 195701 (2002)

\bibitem{johnson99}

  W.-L. Johnson, {\it Mat. Res. Bull.} {\bf 24}, 42 (1999)


\bibitem{schneider}

  S. Schneider, {\it J. Phys. Cond. Matter}
  {\bf 13,} 7723 (2001)

 
\bibitem{sastry03}

  S. Sastry, and C. A. Angell, {\it Nature Materials} {\bf 2,} 739
  (2003)

\bibitem{frenkel} 

  P. R. ten Wolde, and D. Frenkel, {\it Science} {\bf 277,} 1975
  (1997)

\bibitem{oxtoby} 

  V. Talanquer, and D. W. Oxtoby, {\it J. Chem. Phys.} {\bf 109,}
  223 (1998)

\bibitem{sciortino05}

  F. Sciortino, {\it J. Phys. Cond. Matter} {\bf 17,} v7 (2005)

\bibitem{limei06}

  X. Limei, S. V.  Buldyrev, C. A.  Angell, and H. E. Stanley, {\it
    Phys. Rev. E} {\bf 74,} 031108 (2006)

\bibitem{limei09}

  X. Limei, S. V.  Buldyrev, N. Giovannibattista, C. A.  Angell, and
  H. E. Stanley, {\it J. Chem. Phys} {\bf 130,} 054505
  (2009)

\bibitem{doye} 

  J. P. K. Doye, D. J.  Wales, F.  Zetterling, and M. Dzugutov, {\it
    J. Chem. Phys.} {\bf 118,} 2792 (2003)

\bibitem{moriarty} 

J. A. Moriarty,  and M. Widom,  {\it Phys. Rev. B} {\bf 56,}
7905 (1997)

\bibitem{mermin} 

  N. W. Ashcroft, and N. D. Mermin, {\it Solid State Physics, Harcourt
    Brace \& Co } (1976)

\bibitem{hansen}

  J.-P. Hansen, and I. R. McDonald, {\it Theory of simple liquids,
    Academic Press} (2006)

\bibitem{sadoc}

  J.-F. Sadoc, and R. Mosseri, {\it Geometrical frustration, Cambridge
    University Press} (1999).

\bibitem{ponyatovsky03}

  E. G. Ponyatovsky, {\it J. Phys. Cond. Matter} {\bf 15,} 6123
  (2003)


\bibitem{saika}

  I. Saika-Voivod, F. Sciortino, and P. H. Poole, {\it Phys. Rev. E.}
  {\bf 69,} 041503 (2004)


\bibitem{sciortino} 

  F. Sciortino, P. Tartaglia, and E. Zaccarelli, {\it J. Phys. Chem.}
  {\bf 109}, 21942 (2005)

\bibitem{elenius}

   M. Elenius, and M. Dzugutov, {\it J. Phys. Chem.} {\bf 131},
   104502 (2009)

\bibitem{dobson} 

  C. M. Dobson, {\it Nature} {\bf 426}, 884 (2003)

\bibitem{busch2007}

  C. Way, P. Wadhwa, and R. Busch, Acta Materialia {\bf 55} (2007)
  2977

\bibitem{ Acknowledgements} 

  We thank Prof. C. A. Angell for reading the manuscript, comments and
  discussions, and Prof. F. Sciortino for a discussion. We also thank
  the Centre for Parallel Computers (PDC) at the Royal Institute of
  Technology, Stockholm, and the Ekman Consortium for providing
  computer resources.


\end{thebibliography}
\end{document}